\documentclass[letter]{aa}

\usepackage{txfonts}
\usepackage{xspace}
\usepackage{natbib}
\usepackage{graphicx}

\newcommand{\aof}{A\,0535+26\xspace}
\newcommand{\inte}{\textsl{INTEGRAL}\xspace}
\newcommand{\xte}{\textsl{RXTE}\xspace}
\newcommand{\asm}{\textsl{ASM }\xspace}
\newcommand{\pca}{\textsl{PCA}\xspace}
\newcommand{\hexte}{\textsl{HEXTE}\xspace}
\newcommand{\ibisgr}{\textsl{IBIS (ISGRI)}\xspace}
\newcommand{\ibis}{\textsl{IBIS}\xspace}
\newcommand{\jemx}{\textsl{JEM-X }\xspace}

\newcommand{\spi} {\textsl{SPI}\xspace}
\newcommand{\omc}{\textsl{OMC}\xspace}
\newcommand{\suzaku}{\textsl{Suzaku}\xspace}
\newcommand{\rhessi}{\textsl{RHESSI}\xspace}
\newcommand{\ginga}{\textsl{Ginga}\xspace}
\newcommand{\xspec}{\textsl{XSPEC}\xspace}
\newcommand{\ttm}{\textsl{TTM $\&$ HEXE}\xspace}
\newcommand{\osse}{\textsl{OSSE}\xspace}
\newcommand{\cgro}{\textsl{CGRO}\xspace}
\newcommand{\ftools}{\textsl{FTOOLS 6.0.2}\xspace}
\newcommand{\osa}{\textsl{OSA v5.1}\xspace}
\newcommand{\inaf}{\textsl{IFC-INAF}\xspace}

\begin{document}

\title{\aof in the August/September 2005 outburst observed by \xte and \inte}
\author{I.~Caballero\inst{1} \and P.~Kretschmar\inst{2}
\and A.~Santangelo\inst{1} \and R.~Staubert\inst{1} \and D.~Klochkov\inst{1}
\and A.~Camero\inst{3} \and C.~Ferrigno\inst{4} \and M.~H.~Finger\inst{5}
 \and I.~Kreykenbohm\inst{1,6} \and V.~A.~McBride \inst{7}
\and K.~Pottschmidt\inst{8} \and R.~E.~Rothschild \inst{8} \and G.~Sch\"onherr \inst{1}
\and A.~Segreto\inst{4}  \and S.~Suchy \inst{8} \and J~.Wilms\inst{9} \and C.~A.~Wilson\inst{5}}
\offprints{I.~Caballero, \\
\email{isabel@astro.uni-tuebingen.de}} \institute{ Institut f\"ur
Astronomie und Astrophysik, Sand 1, 72076 T\"ubingen, Germany \and
INTEGRAL Science Operations Centre, European Space Astronomy Centre
(ESAC), Apartado 50727, 28080 Madrid, Spain  \and GACE, Instituto de
Ciencias de los Materiales, Universidad de Valencia, PO Box 20085,
46071 Valencia, Spain \and Instituto di Astrofisica Spaziale e Fisica
Cosmica (IASF-INAF), Via La Malfa 153, 90146 Palermo, Italy
\and Universities Space Research Association, NASA Marshall Space Flight
Center, XD12, Huntsville, AL, USA \and INTEGRAL Science Data Centre,
16 Ch.\ d'\'Ecogia, 1290 Versoix, Switzerland \and School of Physics and
Astronomy, Southampton University, Highfield, S017 1BJ, UK \and CASS, University
of California at San Diego, La Jolla, CA 92093-0424 USA  \and Dr. Remeis
Sternwarte Bamberg, Sternwartstr. 7, 96049 Bamberg, Germany   }

\date{Received $<$date$>$; Accepted $<$date$>$ }

\titlerunning{\aof in the August/September 2005 outburst}

\abstract{} { In this Letter we present results from \inte and \xte
observations of the spectral and timing behavior of the High Mass X-ray Binary 
\aof during its August/September 2005 normal (type I) outburst
with an average flux $F_{(5-100)}\,_{\mathrm{keV}}$\,$\sim$400\,mCrab.
The search for cyclotron resonance scattering features (fundamental
and harmonic) is one major focus of the paper. } { Our analysis is
based on data from \inte and \xte Target of Opportunity Observations
performed during the outburst. The pulse period is determined. X-ray
pulse profiles in different energy ranges are analyzed. The broad
band \inte and \xte pulse phase averaged X-ray spectra are studied.
The evolution of the fundamental cyclotron line at different
luminosities is analyzed.} {The pulse period $P$ is measured to be 
103.39315(5)\,s at MJD 53614.5137. Two absorption features are detected in the phase averaged spectra
at $E_{1}\sim$45~keV and $E_{2}\sim$100~keV. These can be interpreted as the fundamental
cyclotron resonance scattering feature and its first harmonic and
therefore the magnetic field can be estimated to be $B\sim$4$\times
10^{12}$\,G.} {}

\keywords{X-rays: binaries - stars:magnetic fields -- stars:individual:\aof }

\maketitle

\section{Introduction}\label{sect:intro}
Discovered in 1975 by \cite{rosenberg75}, the Be/X-ray binary pulsar
\aof \footnote{Referred to as 1A 0535+262 in SIMBAD} consists of an X-ray pulsar with a pulse period of $P$$\sim$103\,s
in an eccentric orbit \citep[$e$$\sim$0.47,][]{finger94_2} of 
$P_{\mathrm{orb}}$$\sim$111\,days \citep{motch91} around the O9.7IIIe optical
companion HDE\,245770 \citep{Li79}. The estimated distance of the
system is $d$$\sim$2\,kpc \citep{steele98}; for a review see
\cite{giovanelli92}. The source was discovered during a giant
outburst (type II), at a luminosity level of
$L_{(3-7)~\mathrm{keV}}$$\sim$$1.2\times 10^{37}\mathrm{erg}\,\mathrm{s}^{-1}$.
Since then, five giant outbursts have been detected: in October 1980
\citep{nagase82}, in June 1983 \citep{sembay90}, in March/April 1989
\citep{makino89}, in February 1994 \citep{finger94}, and in May/June 2005
 \citep{tueller05}. Unfortunately, during the last giant outburst the source
was too close to the sun to be observed by most instruments. It was
only observed by \rhessi \citep{smith05}. However, following the
giant outburst, the source exhibited a normal type I outburst in
August/September 2005, which led to our \inte and \xte Target of Opportunity
Observations. During the peak of the type I outburst, the source showed an
average flux $F_{(3-50)~\mathrm{keV}}$$\sim$$1.9\times
10^{-8}\mathrm{erg}\,\mathrm{cm}^{-2}\,\mathrm{s}^{-1}$  which,
assuming $d$$\sim$2\,kpc, gives $L_{(3-50)~\mathrm{keV}}$$\sim$$
0.9\,\times 10^{37}\,\mathrm{erg}\,\mathrm{s}^{-1}$. Another
 normal outburst took place in December 2005 \citep{finger05}.
In  quiescence, the pulsar behavior appears to be consistent with a
spin-down trend \citep{finger94}. During giant outbursts a spin-up 
has been observed. During the June 1983 giant outburst, a spin-up of
$\dot\nu$$\sim0.6 \times 10^{-11}\,\mathrm{Hz}\,\mathrm{s}^{-1}$
was measured \citep{sembay90}. During the February 1994 giant outburst,
a spin-up of $\dot\nu$$\sim1.2 \times 10^{-11}\,\mathrm{Hz}\,\mathrm{s}^{-1}$
was measured and quasi-periodic oscillations were detected,
confirming the presence of an accretion disk \citep{finger94}.
The X-ray spectrum of the source has been modeled by an
absorbed power law with a high energy cutoff. In the March/April 1989
giant outburst, two cyclotron resonance scattering features were detected
at $E_{1}$$\sim$45\,keV and $E_{2}$$\sim$100\,keV \citep{kend94}. In the February 
1994 outburst, the presence of the fundamental line at $E_{1}$$\sim$45\,keV
was not clear \citep{grove95}. The presence of the fundamental line has been confirmed 
during the August/September 2005 outburst with \inte \citep{kretschmar05}, 
\xte \citep{wilson05} and \suzaku \citep{inoue05} observations.
 
In this paper we report on the analysis of \inte and  \xte observations
of \aof performed during the August/September 2005 outburst.
In Sect.~\ref{sect:data} we describe the observations and data analysis.
In Sect.~\ref{sect:timing} we present the pulse period determination of the source
and pulse profiles in different energy ranges from
$\sim$2\,keV to 200\,keV. In Sect.~\ref{sect:spec} we center
on the analysis of the phase averaged spectra, on the measurement
of cyclotron resonance scattering features and on the evolution of the fundamental line
in different luminosity states. In Sect.~\ref{sect:summ}
we present a summary and conclusions.

\section{Observations and data analysis}\label{sect:data}

\subsection{Instruments: \inte and \xte}\label{sect:instr}
\inte \citep{winkler03} carries two main gamma
ray instruments, the spectrometer \spi \citep[20\,keV--8\,MeV,][]{vedrenne03}
and the imager \ibis \citep[15~keV--10~MeV,][]{ubertini03}, as well as two monitoring
instruments in the X-ray and optical ranges, \jemx \citep[3--35\,keV,][]{lund03}
and \omc \citep{mas03}.

\xte \citep{bradt93} carries three instruments: The
Proportional Counter Array \pca \citep[2--60keV,][]{jahoda96}, the
High Energy X-ray Timing Experiment \hexte \citep[20--200\,keV,][]{rothschild98}
and the All-Sky Monitor \asm \citep[2--10\,keV,][]{levine96}.

\inte data were reduced using \osa. \ibisgr spectral analysis was performed
using alternative calibration files developed at \inaf Palermo
\citep[see][]{mineo06} and a response matrix calibrated with simultaneous
Crab observations, which was in the same field of view as \aof. 
For the spectral analysis we applied a systematic error of
3\% to \textsl{JEM-X}. This has been  evaluated on the basis of the Crab spectrum 
extracted for the same observation as \aof, and agrees with the value suggested by \cite{paizis05}.  
The analysis of \xte data was performed using \ftools.
\vspace*{-3.mm}
\subsection{Observations}\label{sect:obs}

\begin{figure}
  \resizebox{\hsize}{!}{\includegraphics{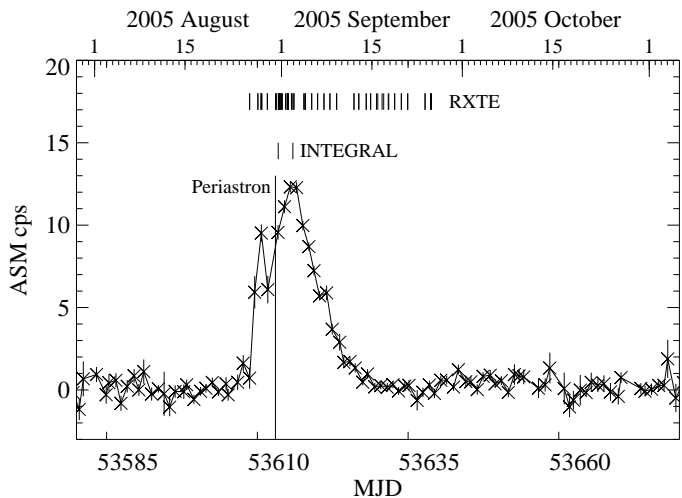}}
\vspace{-4.mm}
  \caption{ \xte \asm light curve of \aof during the normal outburst in
August/September 2005. The start and stop times of our \inte
observation are indicated, as well as each of the individual \xte
observations and the periastron passage.}\label{fig:asm_lc}
\end{figure}

Fig.~\ref{fig:asm_lc} shows the \xte \asm light curve of \aof during
the August/September 2005 outburst. \inte  observations were made close
to the peak with an exposure time of 198.4\,ks. 
The \xte monitoring was performed between the
26th of August and the 24th of September 2005. In this paper we focus on observations 
performed during the peak of the outburst and its decline, for a total exposure time of 125.5\,ks for \xte. Table \ref{tab:obs_log} shows a summary of the observations used in the analysis.\\

\begin{table}
\caption{Summary of \inte and \xte observations used in our
analysis. \xte data are from 37 short observations. The start time
of the first observation and the stop time of the last observation
are given, as well as the total observation time.}
\label{tab:obs_log} \centering
\begin{tabular}{lll}\hline\hline
      Observation          & start time-stop time (MJD)  & duration (ks)     \\\hline
      \inte                                                                        \\
      31 Aug.--2 Sept.2005  &     53613.46--53615.89             &   198.4           \\
      \xte                                                                         \\
      31 Aug.--24 Sept.2005 &     53613.49--53637.84             &   125.5           \\
\hline
\end{tabular}
\end{table}

\vspace*{-0.65cm}

\section{Timing analysis}\label{sect:timing}
\subsection{Pulse Period search}\label{sect:period}
Using epoch folding, we calculated the pulse period of \aof after
barycentering the arrival times and correcting for the orbital
effect. The new ephemeris of \citet{finger06} was used for the binary
correction. To determine $P$ and $\dot{P}$ with high accuracy,  we divided 
the \inte \ibis observation into 27 intervals of $\sim$6\,ks each and folded the
light curve of each of those intervals over the period obtained from
epoch folding. We then performed a phase connection analysis similar
to \cite{ferrigno06}. We found a period of $P$=(103.39315\,$\pm$0.00005)\,s 
for MJD~53614.5137 and a formal (non-significant) value for a spin-up of 
$\dot{P}$=($-3.7$\,$\pm$2.0)\,$\times$~$10^{-9}$~$\mathrm{s}\,\mathrm{s}^{-1}$.
\vspace*{-3.mm}
\subsection{Pulse Profiles}\label{sect:pp}
\begin{figure}
  \resizebox{\hsize}{!}{\includegraphics{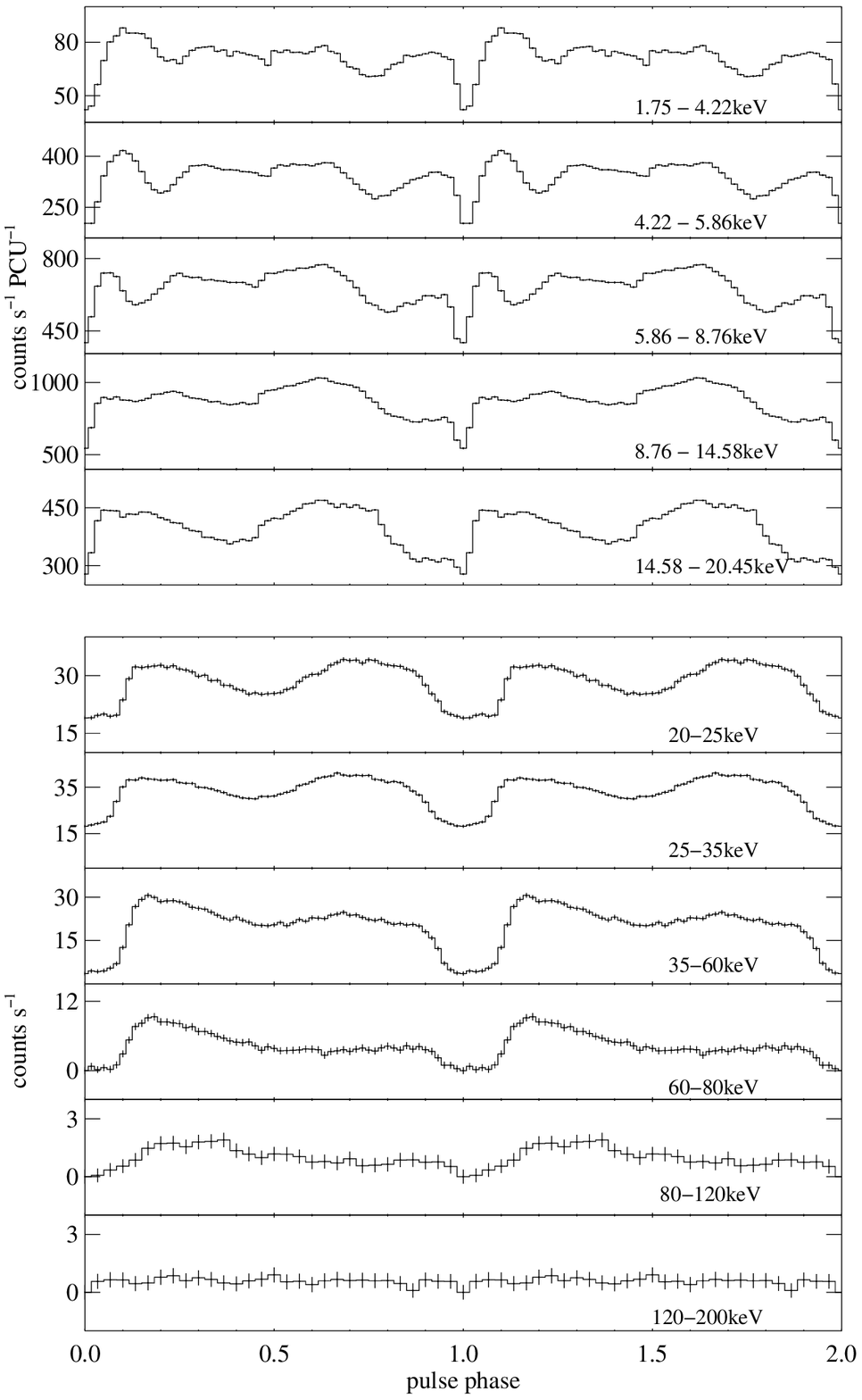}}
\vspace{-20.mm}  \\
  \caption{\pca(top) and \ibisgr (bottom) pulse profiles of \aof.}
  \label{fig:xte_int_pp}
\end{figure}

Using the determined period, we folded \pca and \ibisgr
light curves extracted in different energy ranges. The resulting
pulse profiles, which cover the energy range 1.75--200\,keV, are
presented in Fig.~\ref{fig:xte_int_pp}. Two
pulse cycles are shown for clarity. Similar to other accreting
pulsars, the source shows a complex profile in the low energy range
($\sim$2--20\,keV). A simpler double peak profile is observed from
$\sim$15\,keV to $\sim$\,60\,keV. At higher energies the second peak
is significantly reduced. The source is observed to
pulsate up to $\sim$120\,keV, while above 120\,keV no modulation is
detected. A detailed analysis of the energy dependent morphology of
the pulse profiles is beyond the scope of this paper and will be
presented in a forthcoming paper.
\\

\vspace*{-0.95cm}
\section{Spectral analysis}\label{sect:spec}
\subsection{\inte}\label{sect:inte}
\begin{figure}
  \resizebox{\hsize}{!}{\includegraphics{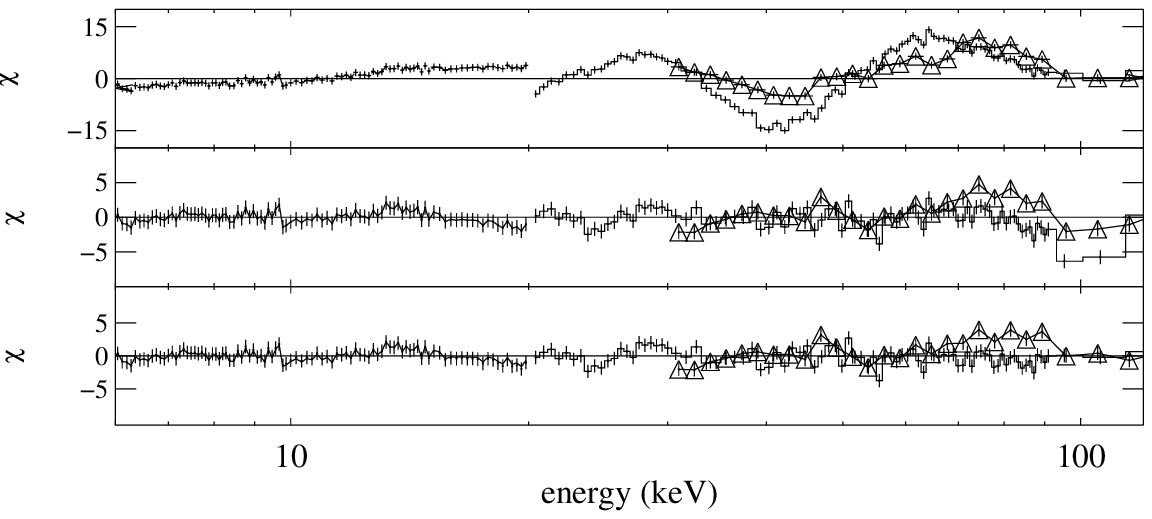}}
\vspace{-18.mm}  \\
  \caption{\inte residuals, corresponding to \jemx (6--20\,keV), \ibis (20--120\,keV) 
and \spi (30--120\,keV, triangles).  
Top panel shows the residuals of a spectral fit without including cyclotron lines
($\chi^2_{\mathrm{red}}=$27.88/218dof). Middle panel shows the
residuals of a fit including one cyclotron line at $\sim$45\,keV
($\chi^2_{\mathrm{red}}=$1.88/215dof). Bottom panel shows the residuals
for a fit including two cyclotron lines at $\sim$45\,keV and
$\sim$100\,keV ($\chi^2_{\mathrm{red}}=$1.37/212dof).}\label{fig:int_res}
\end{figure}

We extracted \inte phase averaged spectra for \emph{JEM-X}, \ibisgr
and \spi. To model the continuum we used a power law with
exponential cutoff (\xspec~ {\tt{cutoffpl}}). We also added to the model an Fe K$\alpha$
fluorescence line at 6.4\,keV and a blackbody component of
$k_\mathrm{B}T$$\sim$1.2\,keV. When fitting this continuum to our
data, two significant absorption-like features are seen in the
residuals at $E_{1}$$\sim$45\,keV and $E_{2}$$\sim$100\,keV (see
Fig.~\ref{fig:int_res}). We modeled these lines using Gaussian lines
in absorption as described in \cite{coburn02}. After the inclusion
of the first line at $E$$\sim$45\,keV,~  $\chi^2_{\mathrm{red}}$
improves from 27.88 (for 218~dof) to $\chi^2_{\mathrm{red}}=$1.88
(for 215~dof). The improvement in the fit when including a second
line is less dramatic. However, including the second line it further
reduces the $\chi^2$  to 1.37 (for 212~dof). The best fit parameters
are reported in Table 2.

\subsection{\xte}\label{sect:xte}
\xte phase averaged spectra for \pca and \hexte data corresponding
to different luminosity states of the source during the outburst
were extracted (Table 2). We modeled the continuum using the same
model as for \inte data, i.e., a power law with an
exponential cutoff, as well as a blackbody component
($k_\mathrm{B}T$$\sim$1.2\,keV) and an Fe K$\alpha$ fluorescence
line at 6.4\,keV. 
For the observations with higher flux and good statistics, two Gaussian 
absorption lines at $E_{1}$$\sim$45\,keV and $E_{2}$$\sim$100\,keV were necessary in the model. 
In some of the observations only one line at $E_{1}$$ \sim$45\,keV was 
included in the model.
For the observations during the decay of the outburst (low luminosity), 
no absorption lines were added to the model.

\begin{figure}
  \resizebox{\hsize}{!}{\includegraphics{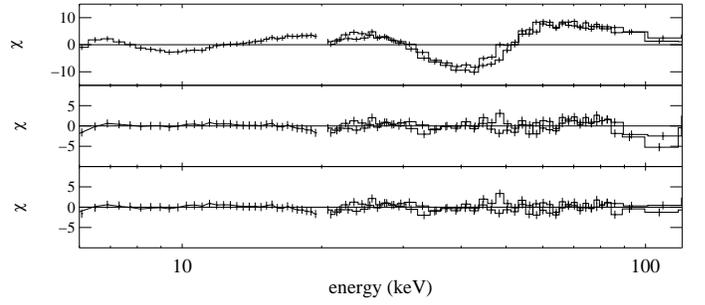}}
\vspace{-18.mm}  \\
  \caption{\xte residuals from observation performed on 1 September 2005.
The upper panel shows the residuals of a fit without
absorption lines ($\chi^2_{\mathrm{red}}=$14.24 for 204~dof). The middle panel shows
the residuals of a fit including one cyclotron line at $\sim$45\,keV
($\chi^2_{\mathrm{red}}=$1.37 for 201~dof). The bottom pannel shows the residuals for a
fit including two cyclotron lines at $\sim$45\,keV and $\sim$100\,keV
($\chi^2_{\mathrm{red}}=$1.03 for 198~dof). }\label{fig:xte_res}
\end{figure}

In Fig.~\ref{fig:xte_res} we show the residuals from fitting
 \xte data from a $\sim12$\,ks observation close to the peak of the outburst, 
performed on 1 September 2005, where the two cyclotron lines were measured 
(see figure's caption for details).

Table \ref{tab:param} shows the best fit parameters
for a selected sample of \xte data. The sample corresponds to observations
around the peak, where two lines are detected,
and at the end of the outburst.

\begin{table*}
\caption{Best fit parameters from simultaneous observations of \inte
  and \xte. \inte data corresponds to observations between 31 August and 2 September, 
close to the peak. The \xte data corresponds to observations performed
on 31 August, 1 and 2 September 2005, and one observation
performed in the declining phase of the outburst on 10 September
2005. The model used for the continuum is the same for all luminosity states.
Uncertainties are 90\% confidence for one parameter of interest
(corresponding to $\chi^{\mathrm{2}}_{\mathrm{min}} + 2.7$).}.
\vspace{-5.mm}
\label{tab:param}
\centering
\renewcommand{\arraystretch}{1.2}
  \begin{tabular}{lcccccccccc}\hline\hline
                 &       $\alpha$                             &
                      $E_{\mathrm {fold}}(\mathrm{keV})$       &  $E_{\mathrm{1}}(\mathrm{keV})$          &
                    $\sigma_{\mathrm{1}}(\mathrm{keV})$       &       $\tau_{\mathrm{1}}$                 &
                    $E_{\mathrm{2}}(\mathrm{keV})$             & $\sigma_{\mathrm{2}}(\mathrm{eV})$        &
                       $\tau_{\mathrm{2}}$                     &        $\chi^{\mathrm{2}}_{\mathrm{red}}/\mathrm{dof}$  \\\hline
\inte            &   $0.54^{+0.05}_{-0.05}$                 &
                $16.7^{+0.4}_{-0.4}$                        &      $45.9^{+0.3}_{-0.3}$             &
                 $~9.5^{+0.3}_{-0.3}$                        &       $0.415^{+0.015}_{-0.015}$          &
                           $102^{+4}_{-3}$               &         $8^{+3}_{-2}$        &
                       $1.1^{+0.4}_{-0.3}$                  &          $1.37 / 212$                                    \\
\xte 31 Aug.   &   $0.59^{+0.14}_{-0.01}$                      & $18.6^{+1.0}_{-0.6}$                      &  
                            $46.7^{+0.6}_{-0.7}$              &
             $10.8^{+0.9}_{-0.8}$                              &      $0.56^{+0.16}_{-0.05}$              &
                     $103^{+6}_{-4}$                     &           $11^{+4}_{-5}$                 &
           $0.9^{+0.4}_{-0.3}$                                 &         $1.12 / 360 $                                     \\
\xte 1 Sept.  &  $0.51^{+0.05}_{-0.03}$                  &
                   $17.3^{+0.3}_{-0.2}$                     &   $45.9^{+0.4}_{-0.3}$                   &
               $10.3^{+0.5}_{-0.4}$                            &    $0.50^{+0.02}_{-0.02}$             &
                   $103^{+3}_{-3}$                       &      $8^{+2}_{-2}$                 &
                $1.1^{+0.4}_{-0.3}$                      &         $1.02 / 360 $                                      \\
\xte 2 Sept.  & $0.51^{+0.02}_{-0.04}$                   &
                     $17.2^{+0.1}_{-0.3}$                   &   $45.5^{+0.4}_{-0.4}$                &
               $10.1^{+0.5}_{-0.4}$                             &  $0.47^{+0.02}_{-0.02}$               &
                     $105^{+5}_{-4}$                     &     $9^{+4}_{-3}$                  &
             $0.7^{+0.2}_{-0.2}$                            &         $1.11 / 360$                                        \\
\xte 10 Sept. &    $0.51^{+0.11}_{-0.07}$                 &
                     $14.0^{+1.4}_{-0.8}$                      &              --                          &
                  --                                           &              --                          &
                          --                                   &              --                          &
                  --                                           &         $0.94 / 372$                                         \\
\hline
\end{tabular}
\renewcommand{\arraystretch}{1.0}
\end{table*}

\begin{figure}
  \resizebox{\hsize}{!}{\includegraphics{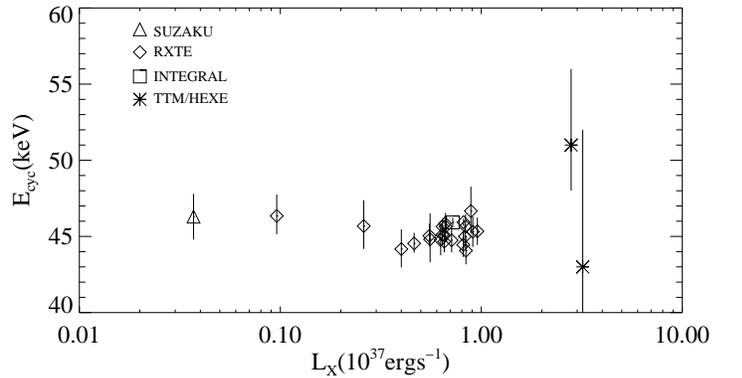}}
\vspace{-8.mm}  \\
  \caption{Evolution of the energy of the fundamental line in
    different luminosity states in the 3--50\,keV range for our
\inte and \xte values. We included data points from \suzaku
\citep{terada06} from the declining phase of the August/September
2005 outburst and values from \emph{TTM/HEXE} \citep{kend94} from
the March/April 1989 giant outburst.} \label{fig:E_lum}
\end{figure}

\section{Summary and Discussion}\label{sect:summ}
Based on \xte and \inte observations of \aof during its
August/September 2005 type I outburst, we detected 
two absorption-like features in the phase averaged
spectrum of the source. These features can be interpreted as electron
cyclotron resonance scattering features (CRSF): a fundamental line
at $E_{1}$$\sim$45\,keV and its first harmonic at $E_{2}$$\sim$100\,keV.
Our findings confirm previous results from \cite{kend94} based on \ttm
observations taken during the March/April 1989 giant outburst and
firmly establish the fundamental line at $E_{1}$$\sim$45\,keV, implying a 
magnetic field of $\sim$4$\times 10^{12}$\,G, from $E_{\mathrm{cyc}}\simeq11.6\,B_{12}$\,keV. 
During the February 1994 giant outburst, observations from \osse on \cgro
clearly show an absorption feature at $\sim$110\,keV, but the
presence of the fundamental line at $\sim$55\,keV was not clear
\citep{grove95}. It was concluded that if the 55\,keV line was
present, its optical depth should have been smaller than that of the
line at 110\,keV, with a ratio $<$1:3.5 (at 95\% confidence).
Analysis of the August/September 2005 outburst shows a different relative
strength of the lines, with the fundamental line deeper than in
the previous giant outburst. The ratio of the depth between the fundamental 
and the first harmonic is $\sim$1:2 for \inte and \xte data.
One possible explanation for these differences could be a different optical depth 
distribution of the electrons in the accretion column, hinting at a different geometry of the
accretion column. Further analysis is ongoing to study the variability
of the line parameters as a function of pulse phase.

We added to our model a single blackbody component with
$k_\mathrm{B}T_\mathrm{bb}\sim1.2$\,keV. Phase dependent blackbody
components with $T_\mathrm{bb}\sim10^{6-7}$\,K have been observed in
the spectra of several accreting pulsars and are generally interpreted
as due to the reprocessing of the hard component by an opaque shell
located at the magnetosphere \citep{oosterbroek97}. However, it
was recently shown by \citet{becker&wolf05} that blackbody-like
components could emerge in the spectra of accreting X-ray pulsars at
low energies. Phase resolved analysis will possibly distinguish
between the two options.

Analysis of accreting pulsars with \ginga \citep{mihara04} and more
recently with \inte \citep{mowlavi06} has shown that the CRSF energy
centroid increases with the decrease of luminosity. This has been
interpreted as due to a variation in the height of the CRSF forming
region. On the other hand, a positive correlation between the energy of the
fundamental line and the luminosity has been observed in GX 301-2
\citep{labarbera05} and recently discovered by \cite{staubert06} in
Her X-1, where it was quantitatively explained with the decreasing
of the height of the line forming region with luminosity when the
source is in the sub-Eddington accretion regime. The centroid energy
of the fundamental CRSF versus luminosity is shown in
Fig.~\ref{fig:E_lum} \cite[d$\sim$$2$\,kpc;][]{steele98}. 
In Fig.~\ref{fig:E_lum} we include values from \suzaku observations of
the declining phase of the outburst \citep{terada06},
as well as values from a \ttm observation of the April 1989 giant
outburst \citep{kend94}. Considering all data from \xte, \inte,
\suzaku and \ttm, no correlation is detected,
 suggesting that the line forming region does not vary with the luminosity 
of the system; further investigation is needed.

\begin{acknowledgements}
We wish to thank the \textsl{INTEGRAL} Mission Scientist, C.
Winkler, and the ESA ISOC personnel for their patient help in
scheduling the observations of this Target of Opportunity Program.
We also thank the ASM/RXTE teams at MIT and at the RXTE SOF
and GOF at NASA's GSFC. 
This research is supported by the Bundesministerium f\"ur Wirtschaft und 
Technologie through the German Space Agency (DLR) under contract no. 50 OR 0302.

Based on observations with INTEGRAL, an ESA project with instruments and science data centre funded by ESA member states (especially the PI countries: Denmark, France, Germany, Italy, Switzerland, Spain), Czech Republic and Poland, and with the participation of Russia and the USA
\end{acknowledgements}
\vspace*{-6.mm}
\bibliographystyle{aa}
\bibliography{ref,a0535}

\end{document}